# The Long and Viscous Road: Uncovering Nuclear Diffusion Barriers in Closed Mitosis

Eder Zavala and Tatiana T. Marquez-Lago*
*Okinawa Institute of Science and Technology, Integrative Systems Biology Unit*
*1919-1 Tancha, Onna son, Okinawa, JAPAN 904-0495*
*e-mail: tatiana.marquez@oist.jp*

Diffusion barriers are effective means for constraining protein lateral exchange in cellular membranes. In *Saccharomyces cerevisiae*, they have been shown to sustain parental identity through asymmetric segregation of ageing factors during closed mitosis. Even though barriers have been extensively studied in the plasma membrane, their identity and organization within the nucleus remains poorly understood. Based on different lines of experimental evidence, we present a model of the composition and structural organization of a nuclear diffusion barrier during anaphase. By means of spatial stochastic simulations, we propose how specialised lipid domains, protein rings, and morphological changes of the nucleus may coordinate to restrict protein exchange between mother and daughter nuclear lobes. We explore distinct, plausible configurations of these diffusion barriers and offer testable predictions regarding their protein exclusion properties and the diffusion regimes they generate. Our model predicts that, while a specialised lipid domain and an immobile protein ring at the bud neck can compartmentalize the nucleus during early anaphase; a specialised lipid domain spanning the elongated bridge between lobes would be entirely sufficient during late anaphase. Our work shows how complex nuclear diffusion barriers in closed mitosis may arise from simple nanoscale biophysical interactions.

## INTRODUCTION

Asymmetric segregation of ageing factors during cell division is essential to maintain parental identity between mother and daughter cells. This is an intense area of research – not only due to its applicability in disease models, but also due to the key role played by asymmetric cell division in the generation of eukaryotic diversity. Cell division is a highly dynamic process, starting at the establishment of polarity between the future mother and daughter cells, continuing via spatiotemporal coordination of lipids and structural proteins involved in membrane remodelling, and ending at cytokinesis.

In contrast to most other unicellular eukaryotes, the yeast *S. cerevisiae* undergoes closed mitosis. That is, its nucleus remains intact at all times, and only breaks down right before cytokinesis. During anaphase, complex changes in the nuclear envelope (NE) result in a dramatic re-shaping of the nucleus: first, the mother lobe buds into a nascent daughter lobe, resembling joined ellipsoids; then, a dumbbell shape emerges, where both lobes remain connected by a long, narrow bridge. As anaphase progresses, cell fate factors become laterally compartmentalized along the cell division axis. Both the rapidly changing nuclear morphology and NE constitution are likely contributors to the compartmentalization of nuclear proteins. As a consequence, discerning their respective contributions has been the focus of much recent research.

By using photobleaching techniques and computational simulations, it was recently shown that geometry changes may account for compartmentalization in the nucleoplasm, while lateral diffusion barriers located between nuclear halves could be responsible for protein segregation at the inner and outer nuclear membranes (INM and ONM, respectively) [1,2]. Even though the nature and structural organization of these diffusion barriers remains elusive, different lines of evidence suggest they depend on specialised lipid domains and scaffolding proteins, such as sphingolipids and septins, respectively.

Sphingolipids are characterized by long, saturated hydrocarbon chains that favour their assembly into tightly-packed, thick bilayers. Sphingolipid-enriched domains are typically more viscous than other lipid phases of the membrane [3], thus effectively reducing molecular diffusion [4]. As membrane proteins have specific affinities for lipid phases depending on their size, amphipathicity, and their membrane anchor, they may become differentially segregated from sphingolipid domains [5]. On a larger scale, morphological changes of the *S. cerevisiae* nucleus during anaphase are compatible with the hypothesis of sphingolipid domains being present at the NE between lobes [6]. This follows from the observation that sphingolipid domains modify the curvature index of membranes [7,8], with a tendency towards negative

---

curvatures such as those found between nuclear halves. Also, recent Fluorescence Loss In Photobleaching (FLIP) experiments suggest that nuclear membrane proteins experience distinct diffusion dynamics at the neck as compared with the lobes [2]. In another line of evidence, the diffusion barrier at the endoplasmic reticulum (ER) slows down the dispersion of membrane proteins therein, thus causing their asymmetric distribution [9]. Moreover, Sur2, a sphinganine hydroxylase necessary for sphingolipid biosynthesis [10], resides at the ER during anaphase [11]. Given the ONM and ER membrane are continuous during this stage [12], the evidences above suggest such membranes' composition at the neck differ from those at the lobes [6]. Thus, distinct diffusion regimes could constitute the underlying protein segregation mechanism preventing free exchange between nuclear halves. However, one should note that irrespective of the presence of sphingolipid domains, other mechanisms such as protein preference for certain types of membrane curvatures [8,13,14] and electric potentials [15], may also play a role in the lateral segregation of nuclear proteins.

Separately, septins are a family of filament-forming, membrane-interacting cytoskeletal GTPases involved in many cellular membrane-remodelling events [16,17]. During mitosis, septin filaments organize into rings and other complex structures [18-20]. They are involved in processes requiring lateral compartmentalization and membrane sculpting into lobular enclosures, as is the case of mitosis and exocytosis [21-23]. In the plasma membrane, septins have been proposed as the main constituents of diffusion barriers [18,22,23]. In addition to their role in hindering protein mobility by forming rings that work as fences, septins possess a lipid-binding motif that enables them to interact with membranes [24]. Hence, it has been proposed that septins recruit and enrich specific phospholipids at the plasma membrane, locally affecting its fluidity [23]. In contrast, while septins have been observed on the inner leaflet of the plasma membrane at the bud neck [19,25], their presence at the ER membrane and ONM is only supported by indirect evidence [1]. An interesting hypothesis is that septin filaments may constrain diffusion at the neck by recruiting and anchoring lipid microdomains [23]. However, it has yet to be shown whether they only recruit the machinery for assembling the barrier, or they are also components of the barrier itself. Notably, these two scenarios are not mutually exclusive, and recently have been referred to as the scaffold model vs. the diffusion-barrier model [25-27].

Lastly, recent experiments suggest that Bud6 [28], a downstream effector of septins, is involved in the formation and maintenance of diffusion barriers that compartmentalize the NE [1] and the contiguous ER membrane [9]. Bud6 stimulates actin nucleation and assembly through the formin proteins Bni1 and Bnr1 during membrane remodelling events [29,30]. In particular, Bnr1 bundles actin filaments [29] and is localized at the bud neck during cell division [31]. In turn, actin polymerization into filaments and other structures is also required for proper nuclear membrane remodelling during anaphase [32]. Moreover, it has been shown that septins promote the assembly of actin filaments into rings [33] and that sphingolipids participate in cytoskeletal organization through actin dynamics during endocytosis [34], as well as in membrane remodelling of other cell types at the dividing neck [35,36]. Taking these facts together, a synergy between sphingolipids and structural proteins at the bud neck emerges as a possible, efficient solution for the compartmentalization through diffusion barriers and simultaneous remodelling of membranes.

In this work, we explore the roles of specialised lipid domains and structural proteins, organized as rings, in establishing nuclear lateral diffusion barriers in *S. cerevisiae* during anaphase. We postulate sphingolipids to form such specialised lipid domains, aligning to experimental evidence [5-8]. However, our analysis is not necessarily limited to them. Based on previous results from fluorescence microscopy techniques [2] and introducing spatial-stochastic simulations, we evaluate the plausibility that these molecular complexes constitute the diffusion barrier. As the nuclear morphology changes dramatically from early to late anaphase (EA and LA, respectively), we studied these stages separately. Accordingly, we developed in silico models and simulated specialised lipid domains and protein rings using realistic nuclear geometries based on experimental measurements. Our results show that, in LA, a specialised lipid domain at the nuclear bridge is enough to compartmentalize the nucleus into different diffusion regimes. Moreover, we explored three different specialised lipid domain configurations in LA and found an optimum agreement with experiments when the domain spans the entire bridge [2]. In contrast, we found that a specialised lipid domain and a protein ring must act together at the neck to constrain diffusion between nuclear lobes in EA. Interestingly, the estimated necessary number of proteins at the ring to constitute the diffusion barrier suggests a polymeric, filamentous fence as the most likely scenario. Altogether, our results suggest that, even though the high viscosity and exclusion properties of specialised lipid domains are probable contributors to the diffusion barrier, additional mechanisms become necessary to fully explain asymmetric segregation. Namely, a protein ring-shaped 'fence' and an elongated nuclear morphology in EA and LA, respectively.

## RESULTS

### Spatial stochastic modelling of nuclear diffusion barriers during anaphase

The compartmentalization effects of nuclear diffusion barriers are known to increase alongside anaphase, and are specific to the nucleoplasm, ONM and INM [1,2]. This was shown to be the case by a combination of FLIP assays and stochastic simulations tracking the fluorescence decay of diffusing marker proteins under continued photobleaching of a small region in the mother lobe (Fig. 1). There, the ratio of the daughter over mother lobe durations for losing 30 % of their initial fluorescence defined the degree of compartmentalization (ºCP), where a higher ratio implies a slower bidirectional transmission of nuclear markers. Thus, the ºCP is inversely proportional to the exchange rate between compartments, and it constitutes an indirect measure of the barrier strength. In this work, we used the FLIP profiles reported in [2] to study the compartmentalization of Nsg1-GFP (ONM marker), GFP-Src1 (INM marker), and the nuclear pore complex (NPC, reported by Nup49-GFP). In all these cases, compartmentalization was not explained by geometry alone (see Figs. 2A and 3A-C in [2]). However, such study utilized a single idealized nuclear geometry, and the variation bounds of numerical experiments were very small, as compared to those of FLIP profiles. So, it remained to be shown whether considering diverse cell geometries would significantly change model predictions, or better fit the data instead. To address this, we first developed realistic sets of 3D geometries in EA and LA from a heterogeneous sample of cell geometries (Fig. S1 and Methods). Then, we used these geometries to carry out spatial-stochastic simulations of FLIP experiments, obtaining their corresponding decay profiles (Movies S1-S4). Our simulations indeed show that considering distinct nuclear geometries may well account for the observed experimental variation bounds in FLIP experiments.

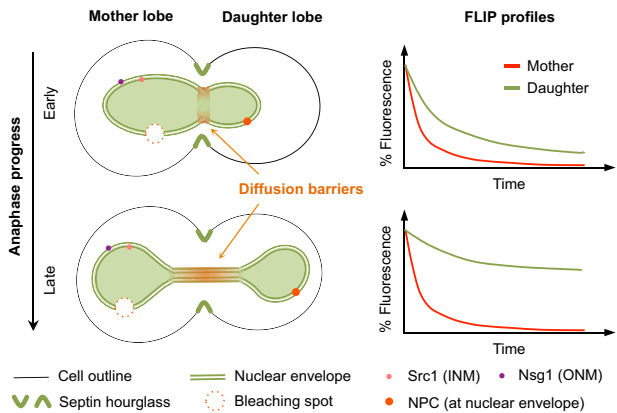

**FIGURE 1. Lateral compartmentalization, assessed by Fluorescence Loss In Photobleaching (FLIP) assays, reflects the diffusion barriers' strength.** The yeast nucleus buds into the daughter cell in early stages of anaphase, elongating into a dumbbell shape in late anaphase. Diffusion barriers have been estimated to locate somewhere between the mother and daughter nuclear halves. Compartmentalization is measured by continuously bleaching the mother lobe while simultaneously measuring the fluorescence decay over time in mother and daughter lobes, separately.

Subsequently, we placed a virtual plane at the neck in EA, and midpoint of the bridge in LA, simulating a hypothetical barrier as in [2]. By fitting to experimental data, we estimated the probability of bidirectional particle transmission and used it as an indicator of barrier permeability. In agreement with previous findings [1,2], our results show that only nuclear re-shaping during anaphase accounts for compartmentalization at the nucleoplasm, whereas diffusion barriers are responsible for compartmentalizing the NE (Fig. S2). Interestingly, the finding that barrier permeability is greater in LA than in EA for NPCs only (Fig. S2) suggests that their compartmentalization is more sensitive to the changing geometry than that of other molecular species.

For the particular case of the NPC, which constitutes the largest diffusing complex in the NE, we further tested whether volume exclusion could generate a crowding effect that hindered its exchange between nuclear lobes. In fact, this effect could easily arise given the narrow thickness of the perinuclear space at the neck and at the bridge in EA and LA nuclei, respectively (Fig. S3). Moreover, given the NPC constituted the largest molecule in our simulations (Table S1), but also the one with the lowest concentration (~150), so far it wasn't clear whether its size would hinder its diffusion at the joint between nuclear halves, thus explaining its compartmentalization. Upon running simulations, we did not find any difference in virtual FLIP profiles when NPC volume exclusion was accounted for or not, in a simplified scenario (Fig. S4). Nevertheless, the fact that many other proteins crowd the membrane did not escape us. However, considering their nanoscale volume exclusion effects is technically impossible at present. Not only are the sizes and diffusion coefficients of most of these crowders unknown, but also there is no guarantee all crowders have been identified already. Moreover, simulating such a huge amount of diffusing particles is computationally prohibitive. Thus, we relied on the previously estimated effective diffusion rates for the reporter proteins used in this study. These rates already account for crowding exerted by the highly inhomogeneous media where proteins diffuse. In addition, previous mathematical models suggest that factors other than excluded volume, such as protein-protein interactions, are contributors to the concentration dependence of lateral mobility [37].

Furthermore, considering time-varying diffusion coefficients didn't improve the fit of our simulations to FLIP data. Specifically, we could not fit our model to experimental observations in EA and LA by assuming a continuous range of varying diffusion coefficients. A proper fit was only possible when assuming independent ranges for EA and LA that would not make biological sense. Hence, the discrepancy found during the first 50 s (Fig. S2) may be related to: a) not considering the fluorophore maturation dynamics; or, most likely b) the fact that the time span of the fluorescence decay profiles reported in [2] is a substantial fraction of the ~500 s duration of the entire anaphase [38], thus representing only a 'snapshot' of a highly dynamic process where the nucleus keeps growing while FLIP experiments take place.

Overall, our spatial-stochastic simulations based on realistic 3D models of a heterogeneous sample of nuclear morphologies supported previous findings, and confirmed that missing variation bounds can be fully accounted for by considering distinct cell volumes away from an idealized average. Moreover, our simulations offer a suitable stage for testing diverse barrier compositions and configurations.

### Specialised lipid domains as diffusion barriers

In contrast to the majority of lipids constituting cellular membranes, the Van der Waals forces between sphingolipid larger backbones, and the associated sterols that stabilize them, result in tighter packing and thicker membranes (Fig. 2). This results in stabilized domains in a liquid ordered ($L_o$) rigid phase with decreased solubility for membrane proteins. Accordingly, sphingolipid domains are good candidates for constituting diffusion barriers by the direct contribution of two effects: an increased viscous drag that slows down protein diffusion within the $L_o$ phase of the domain, and the exclusion of proteins coming from the liquid disordered phase ($L_d$) outside the domain. The latter originates from the hydrophobic matching between the protein amphipathic domains and the membrane where it diffuses. Moreover, in EA, partial depletion of NPCs has been observed at the bud neck; while in LA, loss of fluorescence at the bridge was markedly different from the lobes suggesting different diffusion dynamics (Fig. 3A-C and original images in the Data Viewer, available online in [2]). These observations are compatible with the scenario of sphingolipid domains restricting protein exchange between nuclear halves by hindering their diffusion.

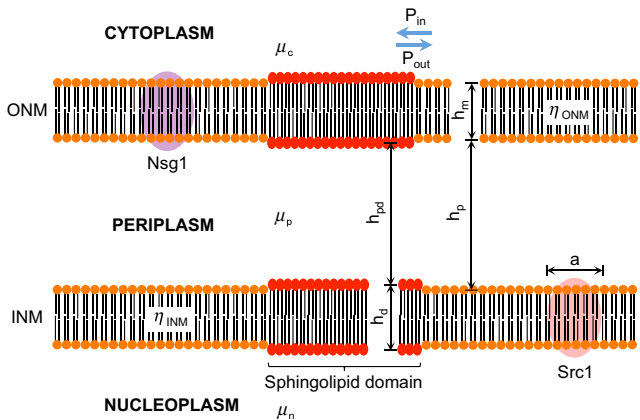

**FIGURE 2. Sphingolipids self-organize into tightly packed, rigid and thicker domains within membranes.** The increased viscosity within these domains causes membrane-bound proteins to diffuse at a lower rate. Moreover, the phase change at the boundary between the domain and the rest of the membrane may work against proteins trying to diffuse into the domain. We call this effect the protein exclusion effect. Measures indicated are: $a$, diameter of membrane inclusions; $h_x$, thickness of diffusive media (membranes and periplasm); $\mu_x$, bulk viscosities (measured in [Pa][s]); $\eta_x$, surface viscosities (measured in [Pa][s][m]); $P_{in}$ and $P_{out}$, probability of proteins diffusing into and out of the sphingolipid domain, respectively.

Following this train of thought, we explored whether specialised lipid domains such as sphingolipid domains account for compartmentalization. To that end, we used the FLIP profiles mentioned above alongside NE dimensions measured by TEM (Fig. S3), and calculated the expected drop in the diffusion rate at the domain by following the Petrov-Schwille model [39] (see Methods). Additionally, we modelled protein exclusion from the domain probabilistically: every time a protein's random walk finds the interphase of a domain when coming from other regions of the membrane, there is a percentage probability $P_{in}$ that it will diffuse into the domain. This probability was estimated by fitting stochastic simulations to FLIP experiments. Conversely, the percentage probability of exiting the domain was fixed at $P_{out} = 100\ \%$, reflecting the preferential protein solubility for ordinary lipids compared to sphingolipid domains.

**TABLE 1. Estimated $P_{in}$ values for membrane proteins in different specialised lipid domain configurations.** Each barrier scenario is graphically described in Fig. 3. For NPC diffusion, we considered both cases where the specialised lipid domain lies only at the ONM or at both the ONM and INM. All values are shown as percentage probabilities. The value $P_{out} = 100\ \%$ was fixed in all cases.

| Protein species | EARLY ANAPHASE | LATE ANAPHASE | | |
|---|---|---|---|---|
| | One ring at neck (300 nm) | One ring at centre of bridge (*300 nm, **100nm) | Multiple rings along bridge | Homogeneous domain spanning the entire bridge |
| Nsg1-GFP (ONM) | 3.5 % | 3.5 % * | 10 % | 15 % |
| GFP-Src1 (INM) | 7 % | 7 % * | 30 % | 35 % |
| NPC (ONM) | 1.6 % | 1.6 % ** | 20 % | 30 % |
| NPC (ONM + INM) | 1.5 % | 1.5 % ** | 20 % | 30 % |

For the domain in EA, we followed observations of reporter proteins delocalization at the bud neck [1] (Fig. 3A and original images in the Data Viewer, available online in [2]), and assumed a 300 nm wide ring shaped domain. Within the domain, we fixed a lower diffusion rate than at other regions of the membrane (Table S2) and estimated $P_{in}$ values by fitting simulations to FLIP data. Importantly, we first verified that a lower diffusion rate at the domain alone (i.e. fixing $P_{in} = P_{out} = 100\ \%$) did not account for compartmentalization (Fig. S5). The estimated $P_{in}$ values for each protein reporter (Fig. S6) are listed in Table 1, where a high $P_{in}$ correlates to lower compartmentalization. As expected, relative $P_{in}$ values reflect the exclusion from the specialised lipid domains experienced by each protein species. It is worth noting NPCs were insensitive to whether the domain is located at the ONM only ($P_{in}$ = 1.6 %) or at both the ONM and INM ($P_{in}$ = 1.5 %).

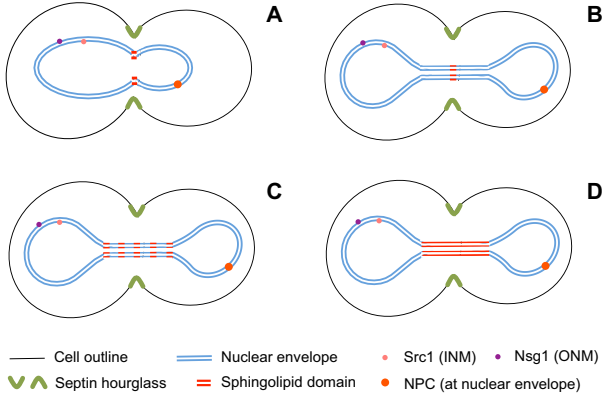

**FIGURE 3. Possible diffusion barrier scenarios, constituted by different specialised lipid domain configurations.** For early anaphase: (**A**) a ring-shaped domain at the neck of mitotic nuclei in early anaphase. For late anaphase: (**B**) a single ring-shaped domain centred at the bridge between nuclear lobes, (**C**) a set of parallel rings uniformly spaced along the bridge, and (**D**) a homogeneously distributed domain spanning the entire bridge length.

We then wondered if, during LA, a single specialised lipid ring domain would account for compartmentalization as it did in EA. To test this, we carried out simulations in LA nuclei, placing the domain at the centre of the bridge connecting the lobes (Fig. 3B). However, this time we fixed $P_{in}$ values to those previously found for EA and estimated the ring's width that would better fit the experimental data (Fig. S7). Surprisingly, a ring domain 300 nm wide fitted the FLIP profiles for Nsg1-GFP and GFP-Src1 in LA best, just as it was the case for EA. For the NPCs, a narrower ring 100 nm wide provided the best fit instead, regardless of whether the domain was assumed to be at the ONM only or at both the ONM and INM. The discrepancy between the rings' width necessary for compartmentalization in LA suggests that additional mechanisms must be accounted for. In what follows, we explore the spatial configuration of domains and the implications that nuclear elongation during anaphase has on them.

Taken together, these results show the higher viscosity and exclusion properties of specialised lipid domains are suitable mechanisms for compartmentalizing nuclear lobes.

## Organization of specialised lipid domains in early and late anaphase

The finding that, in LA, specialised lipid and single-ring domains of different widths compartmentalize our membrane markers is puzzling. Thus, we considered additional domain arrangements, assuming $P_{in}$ reflects lipid-protein interactions causing protein exclusion and no other physical obstacles are present at the diffusion barrier. Accordingly, we developed two additional domain configurations in LA (see Methods). On the one hand, a domain constituted by a series of parallel rings, each 300 nm wide, distributed along the entire bridge length (Fig. 3C); on the other, a continuous domain spanning the entire bridge length (Fig. 3D). As before, we simulated FLIP experiments on these LA nuclei to estimate new effective $P_{in}$ values. The estimations are shown in Fig. S8 and the resulting percentage probabilities $P_{in}$ that showed the best fit are listed in Table 1, where the probabilities for the single ring domain configuration are also shown for comparison. Notably, the estimated $P_{in}$ values for these novel domain configurations are considerably larger than when assuming a single ring domain.

Now, to quantitatively assess the strength of specialised lipid domains in a physically meaningful way, we calculated the transmission coefficient for each scenario in Table 1, and their associated spatial configurations (Fig. 3). This coefficient $\theta_j$, accounts for the permeability of protein $j$ across the barrier, which depends upon its mobility within the domain, its thickness, and the amount of protein available for moving (see Methods). The results are shown in Fig. 4, where we confirm that the diffusion barrier is stronger in the ONM (Nsg1 marker) than in the INM (Src1 marker) irrespective of the domain configuration. Moreover, experimental evidence shows an increase of the barrier strength as anaphase progresses [2], which would imply a reduction on its transmission coefficient. From Fig. 4, we see that this is only compatible with the scenario of a specialised lipid domain spanning the entire bridge length in LA. In general, NPCs are the species most strongly affected by the barrier, but their ability to permeate across it is independent of whether the specialised lipid domain lies at the ONM or at both the ONM and INM.

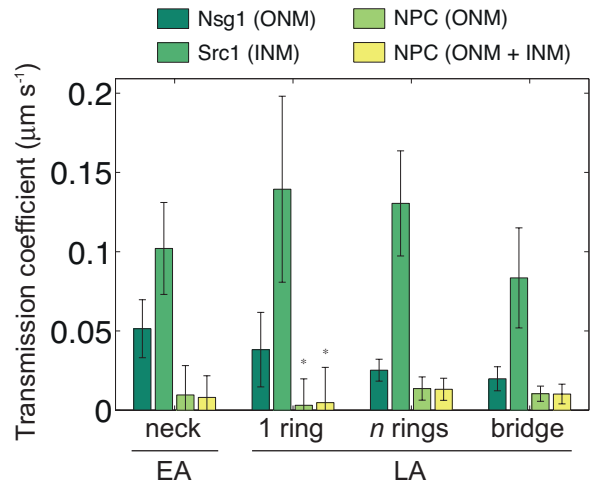

**FIGURE 4. Barrier permeabilities of different specialised lipid domain configurations and for maker proteins at the ONM and INM.** The transmission coefficient of the barrier depends directly on the partition coefficient and the diffusion rate within the domain, and is inversely dependent on its thickness (see Methods). The error bars are associated to the heterogeneous distribution of surface areas (21 nuclei in EA and 34 in LA). For the single ring configuration in LA, a ring of 100 nm wide was assumed for diffusion of NPCs (asterisks). For the NPC, we considered

both scenarios where the domain lies only at the ONM and INM. Effective diffusion coefficients were fixed as in Table S2 and we fixed $P_{out} = 100\%$ in all cases. The corresponding $P_{in}$ values are shown in Table 1.

To determine which domain configuration in LA better reproduces the biological reality, we simulated FLIP experiments as in Fig. 4C in [2] by placing the bleaching spot at different positions along the bridge (see Methods). At each position, we calculated the ºCP and its inverse (ºCP$^{-1}$), and correlated them with the position of the bleaching spot along the bridge relative to the normalized length of the entire nucleus. These experiments are aimed at identifying the position of putative diffusion barriers by observing the intersection of the ºCP and ºCP$^{-1}$ curves. In particular, curves intersecting in a single point would suggest a narrow barrier, whereas curves intersecting at many points (or overlapping) would indicate a distributed barrier. Results from our simulations were compared against FLIP experiments performed in similar conditions [2] and are shown in Fig. 5 for each domain configuration, where the resulting curves from ºCP and (ºCP)$^{-1}$ vs. the bleaching spot position are plotted.

From Fig. 5, one can see that a scenario where the domain is distributed along the entire bridge provides a better fit to the experimental data, when compared to a single central ring. In particular, the homogeneous domain configuration shows a slightly better fit than the multiple rings arrangement. Notably, localization data for Nsg1-GFP shows NPCs are almost completely absent from the bridge during LA (Fig. 3B in [2]), suggesting the diffusion barrier underlying its compartmentalization is also distributed along the entire bridge length.

Overall, our results support specialised lipid domains to be plausible constituents of diffusion barriers, showing a spatial configuration compatible with nuclear morphological changes during anaphase.

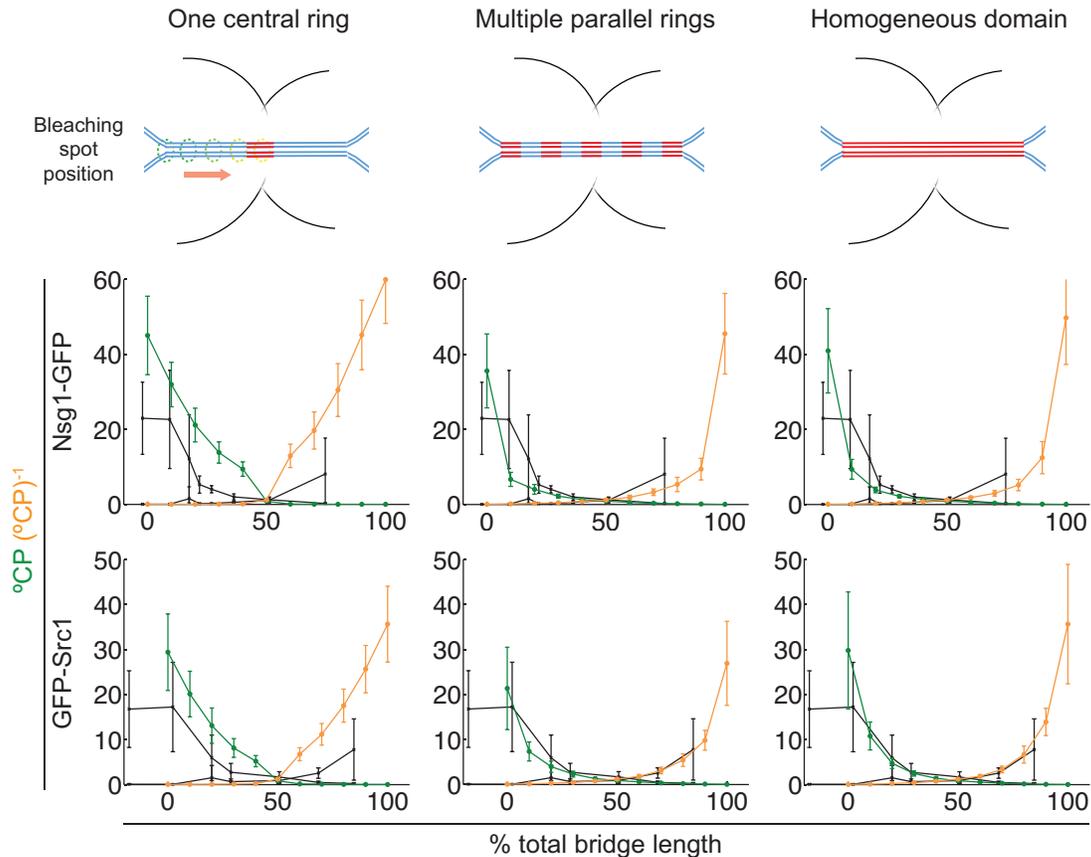

**FIGURE 5. Comparing diffusion barrier scenarios constituted by different specialised lipid domain configurations in LA.** ºCP (green) and ºCP$^{-1}$ (orange) ratios of Nsg1-GFP and GFP-Src1 plotted against the position of the bleaching spot relative to the bridge length (starting at the junction between the mother lobe and the bridge, ending where the latter joins the daughter lobe). $P_{in}$ values were fixed as in Table 1. Experimental data (black) reconstructed from Fig. 4C in [2].

### Protein rings and specialised lipid domains as constituents of the diffusion barrier in early anaphase

When considering a specialised lipid domain organized as one single ring at the neck in EA, the estimated $P_{in}$ values were much lower than those of other domain configurations in LA (Table 1). As $P_{in}$ accounts for protein exclusion effects originated from lipid-protein interactions at the $L_o/L_d$ interphase, and these in turn depend on nanoscale properties that likely remain constant during anaphase, it is highly unlikely that $P_{in}$ would undergo such high variations during anaphase (Table 1) if the diffusion barrier were exclusively constituted of specialised lipid domains. On the other hand, the $P_{in}$ values estimated for EA lie within the same order of magnitude than probabilities of membrane proteins passing through domains corralled by cytoskeleton structures in different cell types [40]. A number of proteins such as septins, Bud6 and actin filament bundles are known to be involved in establishing the diffusion barrier at the plasma membrane of *S. cerevisiae*, but their specific roles in nuclear compartmentalization remain largely unknown. For instance, FLIP experiments revealed a decreased compartmentalization of Nup49-GFP and Nsg1-GFP in the mutant *bud6Δ* [1]. However, whether a regulatory relationship between these nuclear proteins and lipids exists is yet to be seen. So, two very interesting open questions arise: (1) Is the assembly of protein filaments promoted by lipids and membrane curvature? Or, conversely, (2) do protein filamentous structures induce and stabilize nuclear membrane curvature in budding yeast? [17]. Here, it is important to recall that the neck of the dividing nucleus during EA exhibits a large curvature index, and resembles that of the plasma membrane during mitosis. Additionally, evidence from septin organization at the plasma membrane offers a plausible scenario that may also be compatible with the organization of filamentous proteins at the NE [19,24]. In particular, previous studies hint at the possibility that specialised lipid domains may stabilize proteins in an immobile ring configuration [9,22-24,33]. Hence, it is natural to hypothesize that, in addition to the specialised lipid ring domain, a protein ring structure constitutes the diffusion barrier in EA.

To test this hypothesis we assumed that, during EA, the diffusion barrier is constituted by a specialised lipid ring domain and a parallel, immobile protein ring, placed at its centre (see Methods and Movies S1 and S3). This configuration follows from the fact that some filamentous proteins have lipid-binding motifs that contribute to their stabilization within lipid microdomains [23,24]. For simulations, we fixed $P_{in}$ values as in the homogeneous domain in LA (Table 1), and then estimated the necessary number of proteins at the ring to fit FLIP experiments (see Methods). For this scenario, Fig. 6 shows the deviation of our simulations from experiments, where we indicate the numbers of protein at the ring that best fit the FLIP data.

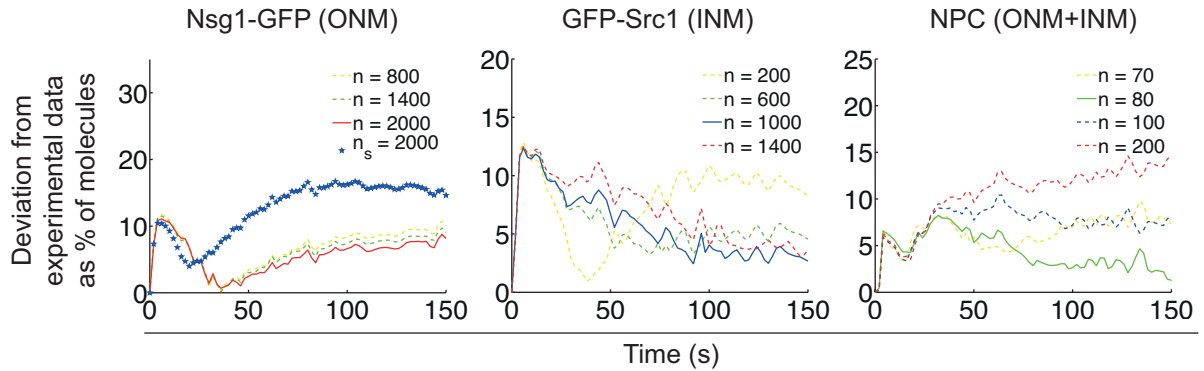

**FIGURE 6. Estimated number n of proteins at the ring constraining lateral diffusion in EA nuclei.** Average deviations (in percentages) of stochastic simulations from the experimental mean for each experimental time step. Mother and daughter lobe deviations are calculated as the absolute value of the difference between simulations and FLIP experiments. The average deviation is the mean of both nuclear lobes' deviations over time. The best fit (i.e. minimal deviation) is shown as a continuous line. For NPC data, we assumed the specialised lipid domain lies at both the ONM and INM, whereas the protein ring is present at the ONM only. Effective diffusion coefficients were fixed as in Table S2 and we fixed $P_{out} = 100\%$ in all cases. $P_{in}$ values were fixed as in the homogeneous domain scenario in LA (Table 1). For Nsg1-GFP, $n_s$ stands for the number of septin proteins in a double ring arrangement.

The small size of deviations estimated in Fig. 6 (in comparison with Figs. S5) suggest that an immobile protein ring embedded at the centre of a specialised lipid ring domain could contribute to the diffusion barrier underlying nuclear compartmentalization in EA. Importantly, we assumed the domain was present at both the INM and ONM, and exhibiting the same biophysical properties in EA and LA. In fact, a scenario where $P_{in}$ and $P_{out}$ values remain constant during anaphase makes more biological sense. This follows from the value of the transmission coefficient associated to the specialised lipid domain alone being $\theta_{Nsg1} = 0.21 \pm 0.04$ µm s$^{-1}$ in EA (recalling $P_{in}$ now takes the same value as in LA), which is ten times higher than its corresponding value in LA (Fig. 4). Assuming the lipid-protein interactions governing the spatial distribution of diffusing proteins remain constant during anaphase, and given that the transmission coefficient is inversely proportional to the barrier thickness $\lambda$ (see Methods), only a tenfold increase in $\lambda$ during anaphase would account for a similar drop in $\theta_{Nsg1}$. Current evidence suggests this is exactly the case since $\lambda \approx 300$ nm in EA [1,2] and the bridge length in LA averages $\lambda \approx 2.85 \pm 0.83$ µm from a heterogeneous sample of 34 nuclei.

The above results suggest that the NE in LA can well be compartmentalized by the combination of a homogeneous specialised lipid domain and an elongated nuclear shape (Fig. 3D and Fig. 5). In contrast, compartmentalization of the EA nucleus can be well achieved by a specialised lipid ring domain and an immobile protein ring acting as a semi-permeable fence, but not by the domain alone. Following this, we have estimated the necessary number of proteins at the ring to account for compartmentalization of Nsg1-GFP, GFP-Src1 and the NPC. Our results show that Nsg1-GFP compartmentalization at the ONM requires double the number of proteins at the ring than GFP-Src1 at the INM does. Moreover, these numbers are rather high ($\sim 10^3$), which suggests a polymeric protein would be a good candidate for constituting the ring. Conversely, NPC compartmentalization at the NE requires much less proteins at the ring ($\sim 80$). Notably, the latter was estimated when we assumed the ring to be exclusively located at the ONM, which follows from evidence suggesting the diffusion barrier is stronger at the ONM than at the INM [2]. Importantly, we also tested whether the high number of proteins at the ring compartmentalizing Nsg1 and Src1 would also compartmentalize NPCs. Assuming $\sim 10^3$ proteins at the ring showed a 30 % to 35 % deviation of the simulations with respect to experimental data. However, a lower number of proteins ($\sim 200$, deviation < 15 % in Fig. 6) already showed signs of a blocked exchange of NPCs between lobes. This could be due to a saturation of the ring at lower threshold concentrations for objects as large as NPCs. As protein ring structures tend to be formed by discontinuous sets of filaments [19], and these in turn are formed by polymerized proteins, it may be that a large fraction of proteins at the ring is contained into filaments. By consequence, a scenario where rather high numbers of proteins at the ring are organized into a small number of filaments of different lengths would similarly compartmentalize large diffusing objects such as NPCs as a small number of non-polymerized proteins would. However, this scenario is not unique and other possible mechanisms are reviewed in the Discussion. For Nsg1-GFP, we followed recent findings regarding septin organization at the plasma membrane [19] and tested whether a septin double-ring (each ring ~4 nm wide and separated ~8 nm from the other) placed at the ONM could as well constrain lateral diffusion (see Methods). We found that, though a protein double ring helps the lipid domains to compartmentalize the ONM, the fit is not better than that of a single ring made of polymeric proteins of a larger size (Fig. 6).

Taken together, these findings suggest that specialised lipid domains likely constitute nuclear diffusion barriers; but also, that the observed compartmentalization arises from a synergistic relationship between such domains and other physical agents. Namely, a protein ring at the nuclear neck and an elongated geometry during early and late stages of anaphase, respectively.

## DISCUSSION

In this work, we focused in compartmentalization of the *S. cerevisiae* nucleus during anaphase. Such compartmentalization is crucial for maintaining parental identity during mitosis, and it has long been questioned whether it is due to diffusion barriers or nuclear geometry changes alone. The study of diffusion barriers in cellular membranes is not only challenging from the experimental perspective, due to several technical constraints, but also theoretically. In such cases, computational simulations offer an important tool for exploring different working hypotheses. This holds specially true when data is scarce, or when studying an organelle that poses difficulties related to single-cell observation and manipulation in real time, such as the nucleus. In that spirit, we addressed the question of how specialised lipid domains and protein rings, two of the most plausible constituents of diffusion barriers, may organize during anaphase to compartmentalize nuclear proteins.

We accounted for morphological changes of the nucleus by studying the early and late stages of anaphase. For this, we developed realistic in silico 3D models from heterogeneous samples of nuclear geometries, and carried out spatial-stochastic simulations with high spatial and temporal resolution. By means of computational modelling, we explored the properties of putative diffusion barriers in each nuclear enclosure and phase, and coupled them in a comprehensive biological picture.

Our first round of simulations confirmed correctness of previous findings in [2]. Notably, this was concluded after carrying out more realistic simulations on a heterogeneous sample of nuclear morphologies, as opposed to an idealized, average geometry. It's important to emphasize that we relied on previously estimated effective diffusion rates for the reporter proteins considered in this study. These rates already account for crowding exerted by the inhomogeneous media where proteins diffuse. However, local variations of diffusion coefficients are perfectly possible when using free diffusion values. In fact, this is exactly what happens to proteins diffusing within membranes populated by lipid microdomains [41,42].

Given that sphingolipid domains are suitable candidates for constituting membrane diffusion barriers, we explored whether their physical properties could account for the observed compartmentalization. Since protein transitions between membrane phases depend on the protein's tertiary structure and amphipathicity, an exact determination of sphingolipid domains' protein exclusion values would require direct measurement of the lipid-protein dynamics at the interphase. Even though these measurements are beyond the scope of this manuscript, we can safely estimate $P_{in}$ and compare its relative value among nuclear enclosures to extract useful information about the overall barrier strength. We did this by calculating the transmission coefficient of the barrier, which quantifies the barrier permeability in a physically meaningful way. Comparing this coefficient for the inner and outer nuclear membrane and during early and late stages of anaphase showed the sphingolipid domain hypothesis is in agreement with experimental evidence.

From the different possible scenarios in which specialised lipid domains could organize in EA and LA, we chose those configurations that better match previous experimental observations. Our results suggest that, in LA, not only is the diffusion barrier present along the entire length of the nuclear bridge, but most likely it is constituted by a homogeneously distributed specialised lipid domain. On the other hand, while we showed that compartmentalization in EA can originate from a specialised lipid ring domain alone, the estimated $P_{in}$ values suggested an additional mechanism must contribute to restrict lateral exchange at this stage. Then, by assuming a protein ring overlapped with the domain, subsequent simulations reproduced the observed compartmentalization between nuclear lobes. Furthermore, we estimated the number of proteins at the ring that were necessary to reproduce experimental FLIP profiles [2]. In what respects to Nsg1-GFP and GFP-Src1 compartmentalization (ONM and INM, respectively), our estimations agree with the scenario of small proteins polymerizing to form stable, immobile ring structures. However, compartmentalization of NPCs required a much lesser amount of proteins at the ring. As the NPC is a much larger diffusing particle than the other markers, and it's diffusion occurs within a more complex environment (ONM + INM + periplasm), other segregating mechanisms may be playing an important role. Constriction of the membrane at the neck, for instance, is likely to require the coordination of anchoring proteins that align the position of the nuclear neck with the mitotic neck of the cell. Thus, it may be that these scaffold proteins aggregate within the specialised lipid domain at the neck and hinder diffusion of NPCs, but not that of other smaller proteins. On the other hand, NPCs remain stable by inducing important membrane deformations in their vicinity [43]. The size of these local deformations (spanning up to ~100 nm) is large enough to significantly affect the way the pore interacts with scaffolding protein rings at the membrane and with the narrow specialised lipid domain at the neck in EA (~300 nm). This may constitute another exclusion mechanism of the barrier since the highly negative curvature index of the nuclear envelope at the neck may severely hinder the ingression of NPCs in the first place [13,14]. Unfortunately, not only are these mechanisms beyond the scope of our model but also, there is a generalized lack of high-resolution experimental studies tracking NPC segregation dynamics at this spatiotemporal scale.

Overall, our results point to a plausible scenario where the diffusion barrier is composed of specialised lipids, but selectively requires additional biophysical mechanisms contributing to it during early and late anaphase stages.

Namely, a protein ring that hinders molecular exchange between mother and daughter lobes in EA, and an elongated nuclear morphology that causes the same effect in LA. Among the proteins reviewed in our introduction that may work as fences, septins are already known to be involved in establishing a NE diffusion barrier [1]. Septins are a known component of the cytoskeleton, providing mechanical support to cellular membranes. During anaphase, an hour-glass shaped, gauze-like septin structure provides support at the neck of the *S. cerevisiae* plasma membrane [19]. It is yet to be seen whether such a similar structure exists at the level of the NE. In fact, this may be the case during EA, when both mother and daughter nuclear lobes have a prolate ellipsoid shape, just as the mother and budding cells that contain them. However, it is very challenging to experimentally assess how currently identified septin structures, as well as the other proteins they recruit, could support the morphological changes of the NE during anaphase. Another open, very interesting question relates to how septins are involved in shaping the junction between nuclear lobes during both anaphase stages and, at the same time, recruit the machinery for sphingolipid biosynthesis. While our manuscript was undergoing final revisions, an interesting study was published reporting that, in wild type anaphase yeast cells, the reduced abundance of NPCs in the NE at the bud neck is dependent on Bud6 and Sur2 [44]. In the same study, staining of lipid species other than sphingolipids was also reduced in the NE at the bud neck and was partially dependent on Bud6 and Sur2 function. These findings are consistent with our model of specialised lipid domains and protein rings as components of the barrier, where sphingolipids and septin-recruited proteins such as Bud6 are good candidates. On the other hand, it is also possible that sphingolipid domains are the sole agents shaping the nuclear bridge in LA, due to the local changes in membrane curvature induced by them [7]. For instance, loss of Spo7, a protein part of a phosphatase complex that represses phospholipid biosynthesis, causes anomalous shaping of the nuclear membrane only in the cytosolic regions, leaving the bridges connecting lobes in LA intact [6]. Experimental evidence will determine whether septins actually constitute a physical obstacle for membrane-bound proteins. However, our simulations suggest this is rather unlikely for LA nuclei. Instead, our study suggests that a homogeneous specialised lipid domain alone may better explain a diffusion barrier spanning the entire bridge length in LA.

Our model involving a protein ring in early, but not in late anaphase, is in agreement with other lines of evidence. Previous works showed that deleting Bud6 or the ring-promoting septin Shs1 [20], which has been shown to decrease the °CP of Nsg1-GFP in EA [1], have no effect on Nsg1-GFP compartmentalization in LA dumbbell nuclei [2]. This implies that, at least in the ONM, the diffusion barrier is regulated differently in EA, as compared to LA. Accordingly, our simulations show that the protein rings' contribution to compartmentalization is required in EA, but not in LA. Moreover, we also found that a lesser-populated ring is required at the INM than at the ONM, suggesting that effects on the former may indirectly arise from a scaffolding protein constituting a ring in the latter. The higher $P_{in}$ estimated for GFP-Src1 (INM), compared to Nsg1-GFP (ONM), may be related to the former being a larger protein than the latter (Table S1). This follows from proteins embedding into membranes according to their size, tertiary structure and amphipathicity, with respect to the membrane thickness [45]. In addition, accumulation of scaffolding proteins at the junction of the lobes may effectively thicken the membrane and increase its exclusion properties [46].

On the other hand, the estimated $P_{in}$ values for the NPC when the specialised lipid domain is assumed to exist only at the ONM or at both ONM and INM are strikingly similar. This suggests that the NPC is less sensitive to whatever barrier may exist at the INM, and that mostly the domain at the ONM (or else, a cluster of proteins working as immobile obstacles) determines its lateral exchange. On the other hand, compartmentalization of the INM has been shown to markedly occur during LA, and its unlikely that it is caused by INM proteins interacting with scaffolding proteins [2]. Thus, there exists the possibility that the NPC, because of its large dimensions, experiences the viscous drag caused by the specialised lipid domain, but not its protein exclusion properties. Hence, its lateral compartmentalization may originate from a slower diffusion rate at the barrier in addition to a blockage caused by protein fences in both stages of anaphase. Recently, however, a novel mechanism was discovered for controlling the redistribution of NPCs during anaphase that is compatible with a model of temporal release of the barrier [47]. Thus, further experiments are necessary to fully understand the complex relationship between dynamical diffusion barriers and its segregating effects on NPCs.

In summary, we propose a plausible model for how diffusion barriers may be constituted and organized in the *S. cerevisiae* nucleus during closed mitosis. The model is based on the biophysical properties of two molecular complexes, sphingolipid domains and protein rings, which are known to be involved in diffusion barriers in other cellular membranes [35,36,48-51]. Importantly, we propose that, while compartmentalization during EA requires a synergy between a specialised lipid domain and a protein ring; the latter is not necessary in LA, where the elongated bridge supersedes this role. This represents a simpler, elegant way *S. cerevisiae* may achieve asymmetrical segregation of ageing factors during closed mitosis. Moreover, our model suggests novel experiments and provides quantitative predictions that may be further tested to better understand diffusion regimes in the nucleus. Additionally, it offers a suitable theoretical framework to explore diffusion barriers in other cellular membranes.

## METHODS

### Definition of model geometries

<u>Nuclear hulls and bleaching spot</u>

Nuclear geometries were developed based on data obtained from fluorescence microscopy (reconstructed from Fig. 2A and Fig. 3A-C in [2] and the original image data available therein) and verified by TEM observations (Fig. S3). The spatial dimensions taken into account are shown in Fig. S1 for EA and LA. In total, 21 nuclei were constructed for EA and 34 for LA.

Geometric functions describing the shapes of nuclear lobes in EA and LA were developed in *Mathematica*. For EA, prolate ellipsoids joined at the tip are good representations of budding nuclei (Fig. S1, A and Movie S1). However, as LA nuclei have a high heterogeneity of nuclear lobe sizes and shapes, we found a Longchamps piriform function to be a better description. This function generates teardrop shaped lobes, and includes a parameter that characterizes deviation from a spherical shape (Fig. S1, B and Movie S2). To represent the nuclear bridge in LA, the lobes were joined together by a cylindrical shape of fixed diameter. Importantly, this process was the same for the ONM and INM, where the only difference was that the perinuclear space thickness was subtracted from the ONM geometry to obtain the one for INM (Fig. S3).

For the virtual FLIP experiments, a bleaching spot, shaped as a cigar, was placed at the nuclear mother lobe. Given the heterogeneous sizes of nuclei in our study, the absolute position of the bleaching spot varied from cell to cell. However, we followed the criterion used in the photobleaching protocol in [2] and located the bleaching spot at a lateral edge of the mother lobe diameter, centred right on the NE (Fig. 1 and Movies S1 and S2). The geometric 3D models were then transformed into Delaunay triangulations and imported as suitable files to be used in *Smoldyn*.

<u>Domain configurations in late anaphase</u>

For the sphingolipid domain configuration in Fig. 3C, we assumed a series of parallel sphingolipid ring domains, each 300 nm wide, uniformly distributed along the entire bridge length. For the computational simulations, an average bridge length of $2.85 \pm 0.83$ μm (34 cells) (Fig. S2, D in [2]) allowed us to place several 300 nm wide rings along the bridge. In most cases, we allocated five rings (one at the centre, two at the edges connecting with lobes, and the other two equally spaced in between). For cells with the largest bridges (3 cells), we were able to accommodate up to seven rings. For the shortest bridges (6 cells), we could only place three rings. As we assumed equally sized rings, but the bridge length varies from cell to cell, the distance between neighbouring rings could not be fixed. To maintain homogeneous spacing between rings relative to each nucleus' bridge length, such distance was chosen to be never below half or above twice the ring width (300 nm).

In contrast, for the scenario in Fig. 3D, we assumed the barrier was constituted by a homogeneous sphingolipid domain spanning the whole bridge in LA. The borders of this domain were placed at the edges of the bridge, just where it joins the nuclear lobes (Movie S3).

### Spatial stochastic simulations

<u>Model parameters and settings</u>

The particle numbers used for all spatial-stochastic simulations were obtained from quantification data from the Yeast GFP Fusion Localization Database [11,52]. The relative amounts for TetR, Nsg1, Src1 and Nup49 (NPC) are listed in Table S1 and follow those used in [2]. In the case of Nup49, one has to consider that it is located in the inner double rings of the NPC, with 16 Nup49 proteins each [53].

To estimate the rate of the bleaching reaction occurring at the bleaching spot, we relied upon the TetR-GFP diffusion rate already estimated in [2] by means of Fluorescence Correlation Spectroscopy. As TetR-GFP is not compartmentalized in EA, it can be used to estimate the bleaching rate by fitting simulations to the FLIP experiments performed on it. In these simulations, 5,000 TetR-GFP particles were uniformly distributed in the nucleoplasm and diffused at a rate of 1.9 µm$^2$/s. Conversely to a previous estimation [2], we defined the bleaching rate as an irreversible conversion rate for molecules entering the bleaching spot, but allowing particles entering to it to potentially escape without being bleached. This more realistic definition allowed us to estimate the bleaching rate at $k_b = 150$ s$^{-1}$. Notably, this phenomenon is intrinsically related to the laser photobleaching effect on the GFP fluorophore, thus being independent on the nuclear compartment where the particles diffuse (nucleoplasm, ONM or INM). Accordingly, the same bleaching rate was used for all of our FLIP stochastic simulations.

For simulating the FLIP experiments shown in Fig. 5, the bleaching spot was located at different relative positions along the nuclear bridge in LA. This was readily done by dividing the bridge length within each nucleus in ten parts, thus yielding eleven evenly-spaced positions, labelled from 0 % to 100 % of the total bridge length. Importantly, the 0 % position was coincidental with the mother lobe junction with the bridge, while the 100 % was coincidental with the junction of the latter with the daughter lobe. In this way, we were able to compare our simulations with the ºCP and (ºCP$^{-1}$) values obtained previously in Fig. 4C in [2].

Simulating protein rings at the neck of EA nuclei required a few considerations. Namely, that each protein was defined as a hard sphere, thus allowing for volume exclusion among them and within the ring. However, to reflect their polymerization properties in a more realistic way, we allowed a slight overlapping between them, which amounted to ~15 % of their total diameter. The size of the proteins constituting the ring was fixed to 11 nm, comparable to the thickness of septin and acting filaments (~10 nm) [23,54] and Bud6 diameter (~11.6 nm) [55] in *S. cerevisiae*.

Spatial stochastic simulation settings

Stochastic simulations were carried out using *Smoldyn*, a computer program for simulating off-lattice, spatial stochastic chemical kinetics, on a microscopic size scale [56]. In our simulations, particle diffusion was either confined to a volumetric enclosure (TetR-GFP in the nucleoplasm, NPC in the perinuclear space) or surface enclosure (Nsg1-GFP and GFP-Src1 in the ONM and INM, respectively). The effective diffusion rates used in our simulations were estimated previously in [2] (Table S1). Given that *Smoldyn* algorithms are an implementation of Smoluchowski diffusion theory, we are required to provide suitable simulation parameters to achieve accuracy within reasonable spatiotemporal resolution. To that end, we set a time step of 40 µs, which corresponds to a spatial resolution of ~7 nm. This time step was calculated according to $\Delta t \leq s^2/2nD_{max}$, where $D_{max}$ is the fastest diffusing species, $s$ is the desired spatial resolution, and $n$ is the degrees of freedom ($n = 2$ for membrane bound proteins such as Nsg1 and Src1 and $n = 3$ for proteins diffusing within volumetric enclosures). Notably, simulations performed with smaller time step lengths, i.e. higher spatial resolution, didn't yield results significantly different from the ones shown here. The choice of the time step described above allowed us to achieve highly accurate results within a reasonable computational simulation time. Depending on the particular experimental setup (e.g. number of particles, number of interactions, anaphase stage, the complexity of the diffusion barrier and the time span of the FLIP profile), the simulations lasted from 1 to up to 7 days in a HPC cluster for batches of 21 distinct nuclei for EA and 34 for LA.

## Estimating effective diffusion rates at sphingolipid domains

The Saffman-Delbrück model [4] supports the assumption of a decreased diffusion coefficient at the neck preventing free lateral mobility of membrane-bound proteins between the mother and daughter nuclear lobes. This model states that, for the typical scenario found in biological membranes, diffusion coefficients of membrane proteins depend mostly on the membrane thickness and viscosity, rather than in the size of the diffusing particle. The model is given by

$$D_{SD} = \frac{k_B T}{4\pi \mu_m h} \left[ \ln\left(\frac{\mu_m h}{\mu a}\right) - \zeta \right] \quad (1)$$

where the diffusion rate $D_{SD}$ of a cylindrical inclusion of radius $a$, in a membrane with thickness $h$, is determined by the bulk viscosities $\mu_m$ and $\mu$ of the membrane material and surrounding fluid, respectively. A logarithmic law to which the Euler-Mascheroni constant $\zeta \approx 0.577215$ is subtracted governs the diffusion rate dependence on viscosities and particle-to-membrane dimensions. However, the Saffman-Delbrück model is valid only for membrane inclusions that are small compared to the characteristic length scale brought about by hydrodynamics. This hydrodynamic length scale is determined by the ratio

$$l = \frac{\eta_m}{\mu_1 + \mu_2} \quad (2)$$

where $\eta_m = \mu_m h$ is the surface viscosity of the membrane, measured in [Pa][s][m], and $\mu_1$, $\mu_2$ are the bulk viscosities of the fluid flanking each side of the membrane, measured in [Pa][s]. In equation (2), we can assume that the fluids surrounding both sides of the membrane have bulk viscosities equal to that of cytoplasm ($\mu_1 = \mu_2 = \mu_c$). As we want to test diffusion in the ONM, INM and the whole NE, we will also assume that the nucleoplasm and periplasm have the same bulk viscosities than cytoplasm ($\mu_c = \mu_n = \mu_p$). To characterize the ratio of the membrane inclusion to the hydrodynamic length scale we use the non-dimensional reduced radius $\varepsilon$, which is given by

$$\varepsilon = \frac{a}{l} = a\frac{2\mu_c}{\eta_m} \quad (3)$$

Thus, the Saffman-Delbrück model in equation (1) is valid on the condition that $\varepsilon \ll 1$. Even though we can estimate $\mu_c$, we don't know a priori the value of $\eta_m$ for the nuclear membranes. It is likely that for the small sizes of Nsg1-GFP and GFP-Src1 (Table S1), the Saffman-Delbrück model is still valid [57], but we cannot assert the same for the NPC. Thus, we must retort to using a hydrodynamic model describing the mobility of a membrane inclusion of an arbitrary radius for arbitrary viscosities. This is readily available in the Petrov-Schwille generalization of the Saffman-Delbrück model, which is an approximation of an exact model developed earlier [58] and valid for a very wide range of values ($10^{-3} \leq \varepsilon \leq 10^5$) with a relative error below 0.015 % with respect to the exact solution [39]. The Petrov-Schwille model is given by

$$D_{PS} = \frac{k_B T}{4\pi \eta_m} \left[ \frac{\ln(2/\varepsilon) - \zeta + 4\varepsilon/\pi - (\varepsilon^2/2)\ln(2/\varepsilon)}{1 - (\varepsilon^3/\pi)\ln(2/\varepsilon) + c_1\varepsilon^{b_1}/(1 + c_2\varepsilon^{b_2})} \right] \quad (4)$$

where the parameters $c_1 = 0.73761$, $b_1 = 2.74819$, $c_2 = 0.52119$ and $b_2 = 0.61465$ were estimated by Petrov and Schwille to fit the exact solution [39].

From our TEM image analysis (Fig. S3), we estimate a membrane thickness of $\eta_m \approx 4$ nm for both ONM and INM. Topographic data from AFM of plasma membranes populated by sphingolipid rafts estimate up to a ~7 Å increase in the membrane height where the domains are located [59-61]. For a lipid bilayer, this implies the membrane thickens up to $\eta_d \approx 5.4$ nm at the domains. On the other hand, a set of experiments using optical traps to track raft-associated proteins diffusing in the plasma membrane of mammalian cells estimate that they experience a three-fold higher viscous drag than non-raft proteins [3]. Starting from the effective diffusion rates previously estimated for the GFP-tagged proteins mentioned in this study [2], is straightforward to calculate the drop in the diffusion rate at the sphingolipid domain for Nsg1-GFP and GFP-Src1 by using the Einstein-Smoluchowski relation

$$D = \upsilon k_B T \quad (5)$$

where the viscous drag $\gamma$ is the inverse of the mobility $\upsilon$. Thus, a drop to one third of the estimated effective diffusion rate is expected for a three-fold increase in the viscous drag at the sphingolipid domain. However, the scenario is not so simple for the NPC as it diffuses while embedded in the whole NE, which comprises three phases with different viscosities. Notably, the work of Pralle et al. [3] is the only available reference related to direct measurements of viscous drag of non-raft vs. raft-associated proteins. These experiments didn't address more complicated scenarios, as is the case for the NPC. Here, we approximated the viscous drag experienced by the NPC by using the Petrov-Schwille model and combining equations (4) and (5) into:

$$\gamma = 4\pi\eta_m \left[ \frac{\ln(2/\varepsilon) - \zeta + 4\varepsilon/\pi - (\varepsilon^2/2)\ln(2/\varepsilon)}{1 - (\varepsilon^3/\pi)\ln(2/\varepsilon) + c_1\varepsilon^{b_1}/(1 + c_2\varepsilon^{b_2})} \right]^{-1} \quad (6)$$

On the other hand, the bulk viscosity of the cytoplasm has been estimated to be $\mu_c \sim 1.5\mu_w$ [62], where $\mu_w$ is the bulk viscosity of water. At T = 30 °C = 303.15 K, the temperature at which the FLIP experiments were carried out [2], this viscosity is $\mu_w$ = 7.978 x $10^{-4}$ Pa s. Thus, the bulk viscosity of cytoplasm is $\mu_c$ = 11.967 x $10^{-4}$ Pa s. Given that we know the effective diffusion coefficients and sizes of the membrane-bound proteins diffusing in the membrane (Table S1), we can estimate by means of equation (4) the surface viscosities of the ONM, INM and the whole NE (ONM + INM + perinuclear space) experienced by Nsg1-GFP, GFP-Src1 and the NPC, respectively.

In summary, we estimated the drop in the diffusion rate for the NPC at the sphingolipid domain by calculating the viscosities at the INM, ONM and periplasm using the Petrov-Schwille model. Estimations of all aforementioned parameters are listed in Table S2. As expected, the drop of the NPC diffusion rate at the sphingolipid domain is more than three-fold, as it was the case for Nsg1-GFP and GFP-Src1. From equation (6), we can calculate the viscous drag experienced by Nsg1 and Src1 proteins (Table S2). This drag is one order of magnitude larger than measurements carried out before [3]. There are a number of possible explanations for this discrepancy:

1) The plasma membrane components in mammalian cells (where viscous drag measurements were carried out) may be radically different than those of the nuclear membrane of yeast cells. Moreover, in our case of interest, yeast cells are dividing. This is particularly important given that during anaphase cytoskeleton undergoes a dramatic rearrangement that may prevent membrane-bound proteins to diffuse normally.
2) The viscous drag measurements by Pralle et al. [3] were carried out at a temperature of ~36 °C, while the FLIP experiments in yeast used to estimate an effective diffusion rate were carried out at ~30 °C. It is known that viscosity increases with lower temperatures, but it is difficult to assess the magnitude of this effect between different cell types.
3) As the viscous drag is calculated from an "effective" diffusion rate and this rate is estimated from FLIP experiments, it turns out we're actually looking at the sum of many processes affecting diffusion. In other words, not only the presence of sphingolipid domains, but also of membrane proteins constituting physical obstacles at the barrier may be the cause of an over-estimated viscous drag.

### Estimating transmission coefficients

The permeability of a diffusion barrier with respect to a diffusing molecular species $j$ can be accounted by its transmission coefficient $\theta_j$ (also known as the permeability coefficient [63,64]). This coefficient is given by

$$\theta_j = \frac{KD_j}{\lambda} \quad (7)$$

where $K$ is the partition coefficient, $D_j$ is the diffusion rate of protein $j$ within the barrier, and $\lambda$ is the barrier thickness.

The partition coefficient $K$ reflects the distribution of the diffusing protein inside and outside the specialised lipid domain. This can be calculated from the definition of chemical potential

$$\mu_j = \mu_j^0 + RT \ln C_j \quad (8)$$

where $\mu_j^0$ is the chemical potential of protein $j$ in the standard state and $C_j$ is its concentration. In our simulations, and before we set the bleaching reaction to start, the proteins diffusing in the specialised lipid domain phase reach a chemical equilibrium with the proteins diffusing outside the domain (i.e. the net exchange between phases is zero). Thus, its chemical potential $\mu_j$ is the same in both phases, and we found the partition coefficient of the protein is given by

$$K_{in/out} \equiv \frac{C_j(in)}{C_j(out)} = e^{[\mu_j^0(out) - \mu_j^0(in)]/RT} \quad (9)$$

Importantly, in equation (8) we have ignored electrical and other less significant sources of work. While the electrical potential becomes important for diffusing ions and molecules with a high dipolar moment, it can be ignored for the case of uncharged proteins diffusing within a lipid membrane. The term in the right hand side of equation (9) depends on the Gibbs free energy to transfer the protein from one membrane lipid phase to the other, which in turn depends on how energetically favourable is the interaction of the protein with its surroundings. The details on these protein-lipid interactions are beyond the scope of our work, but we can get a fair estimation of the partition coefficient by calculating the ratio of protein surface concentrations inside and outside the specialised lipid domain.

After computing the surface areas of our distribution of virtual nuclear hulls in EA and LA, and counting the number of proteins inside and outside the domain in equilibrium, we calculated $K_{in/out}$ for each of our in silico experiments where a specialised lipid domain was the sole component of the barrier. Taking into account the diffusion rates within the domain (Table S2) and the geometry of the barrier (Fig. 3), we used equation (7) to calculate the transmission coefficients $\theta_j$ shown in Fig. 4.

### Strains and growth conditions

For TEM analysis, yeast strain YYB5528 (WT) [2] was used. Single colonies from freshly streaked plates were incubated into 3 mL YPD media and grown overnight at 30 °C. Cultures were diluted into 3 mL fresh YPD media and grown to OD600 (optical density at 600 nm) of ~1.0. Cells were then prepared for electron microscopy following the protocol in [65].

## ACKNOWLEDGMENTS

The authors wish to thank Zach Hensel and Toshio Sasaki for assistance on yeast culture preparation, fixation and analysis by TEM at OIST. Special thanks to Barbara Boettcher and Zach Hensel for helpful preliminary discussions on the project, and to the Barral lab for yeast strain YYB5528 (WT). Lastly, both authors would like to thank André Leier for assistance with the computational setup in the OIST HPC cluster facility.

## AUTHOR CONTRIBUTIONS

T.M.L. and E.Z. conceived the study, designed experiments, analysed data and discussed results. E.Z. performed numerical simulations and TEM analysis. Both authors wrote the paper and approved the final manuscript.

## SUPPLEMENTARY FIGURES

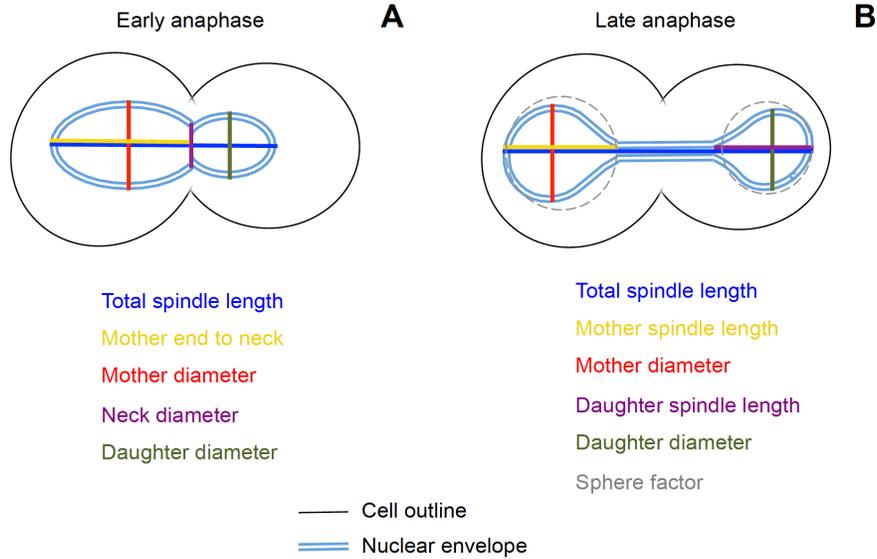

**FIGURE S1. Defining nuclear geometries for simulations.** Spatial dimensions used for developing realistic *in silico* 3D models of yeast nuclei in (**A**) early and (**B**) late anaphase.

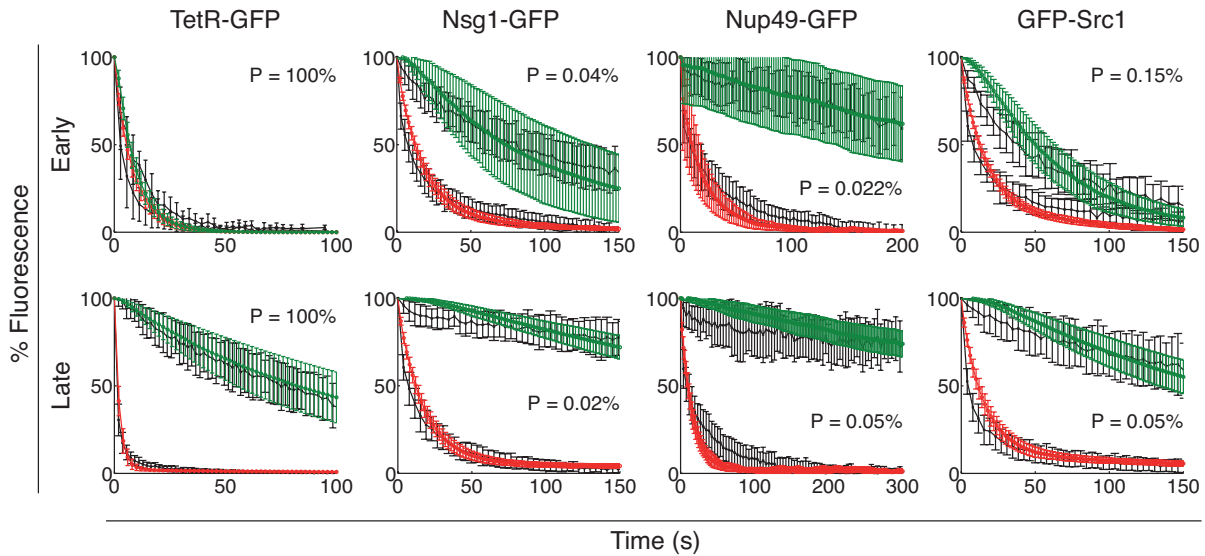

**FIGURE S2. Diffusion barriers compartmentalize nuclear membranes, but not the nucleoplasm in early and late anaphase.** Mean ± SD fluorescence versus time of nuclei in early (21 cells) and late anaphase (34 cells). FLIP experiments (black) are compared with simulations (red for mother and green for daughter lobe). A hypothetical diffusion barrier permeability P was estimated for TetR-GFP (nucleoplasm), Nsg1-GFP (ONM), Nup49-GFP (NE) and GFP-Src1 (INM).

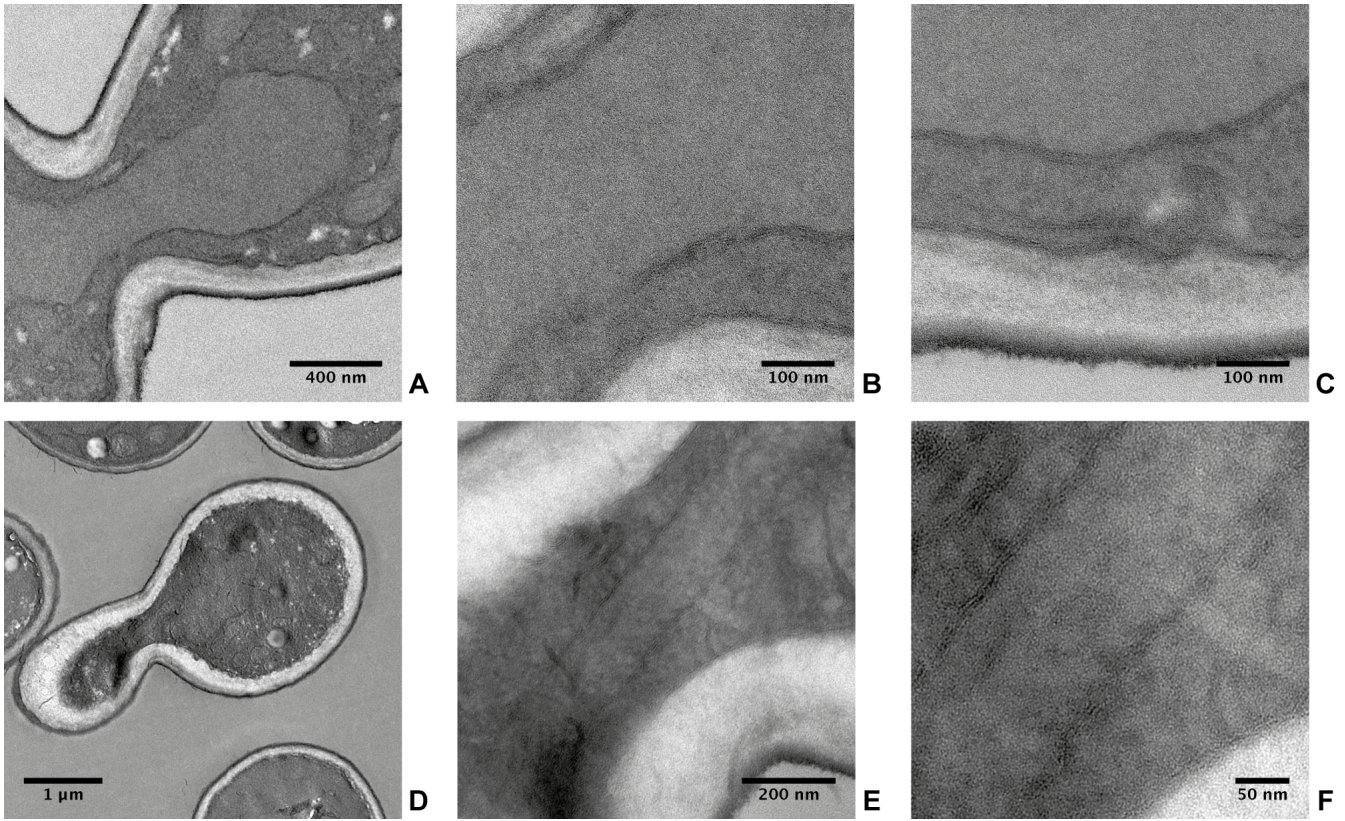

**FIGURE S3. TEM images of yeast nuclei during anaphase.** Each row, from top to bottom, shows cells in (**A, B, C**) early anaphase and (**D, E, F**) late anaphase. Zoomed areas from left column are shown in centre and right columns: early anaphase: (**B**) bud neck and (**C**) daughter nuclear lobe; late anaphase: (**E**) whole bridge and (**F**) bridge at neck. The average thickness of perinuclear space ($h_p$ in Fig. 1, measured between the phospholipid heads of inner lipid leaflets facing the periplasm) was 22 ± 6 nm at nuclear lobes (regardless of the mitotic stage) and 13 ± 4 nm at the connecting bridge (in late anaphase). The staining protocol used was the same as in [65].

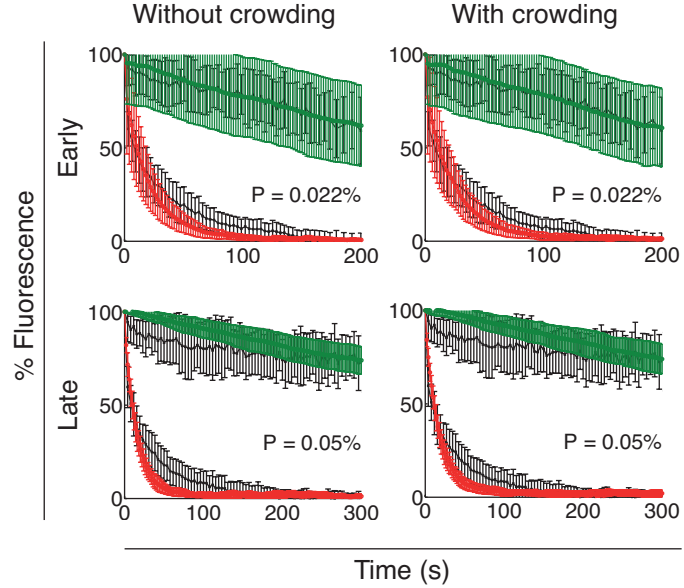

**FIGURE S4. Crowding effects occasioned by volume exclusion of the NPC do not affect its compartmentalization.** Nup49-GFP mean ± SD fluorescence versus time of nuclei in early (21 cells) and late anaphase (34 cells). NPCs were considered as diffusing non-overlapping spheres. Given the number and size of NPCs (Table S1), we estimate a 6.19 % and a 6.55 % crowding of the NE surface in early and late anaphase, respectively. FLIP experiments (black) are compared with simulations (red for mother and green for daughter lobe). The diffusion barrier permeability was fixed at P = 0.022 % for early, and P = 0.05 % for late anaphase, as estimated previously (Fig. S2).

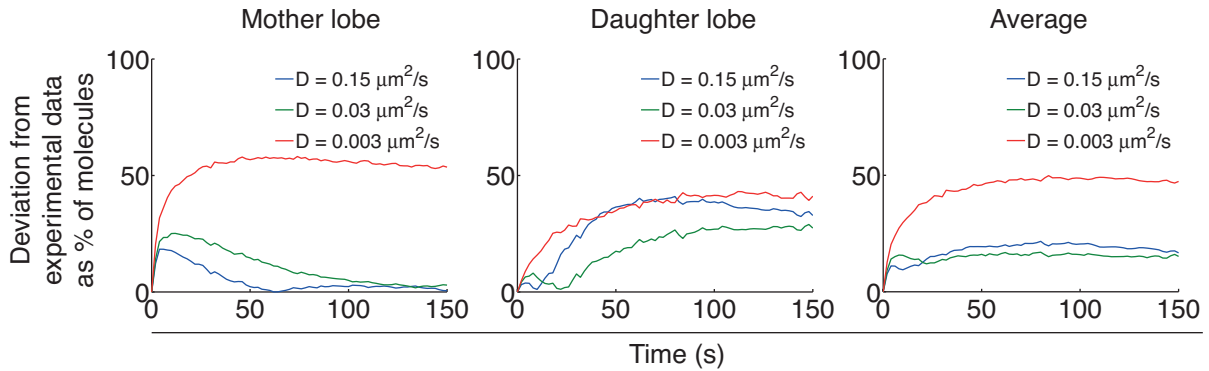

**FIGURE S5. The specialised lipid domain requires protein exclusion together with a decreased diffusion rate.** Deviations (in percentages) of stochastic simulations from the experimental mean, for each experimental time step. Mother and daughter lobe deviations were calculated as the absolute value of the difference between simulations and FLIP experiments. The average deviation is the mean of both nuclear lobes' deviations over time. Data from Nsg1-GFP (ONM) FLIP profiles in EA was used (21 cells). Given the diffusion rate at the lobes was estimated to be $D_{Nsg1-GFP} = 0.3$ μm$^2$/s at the lobes, we explored decreasing that value to a half, a tenth, and a hundredth at the specialised lipid domain; while fixing $P_{in} = P_{out} = 100$ %.

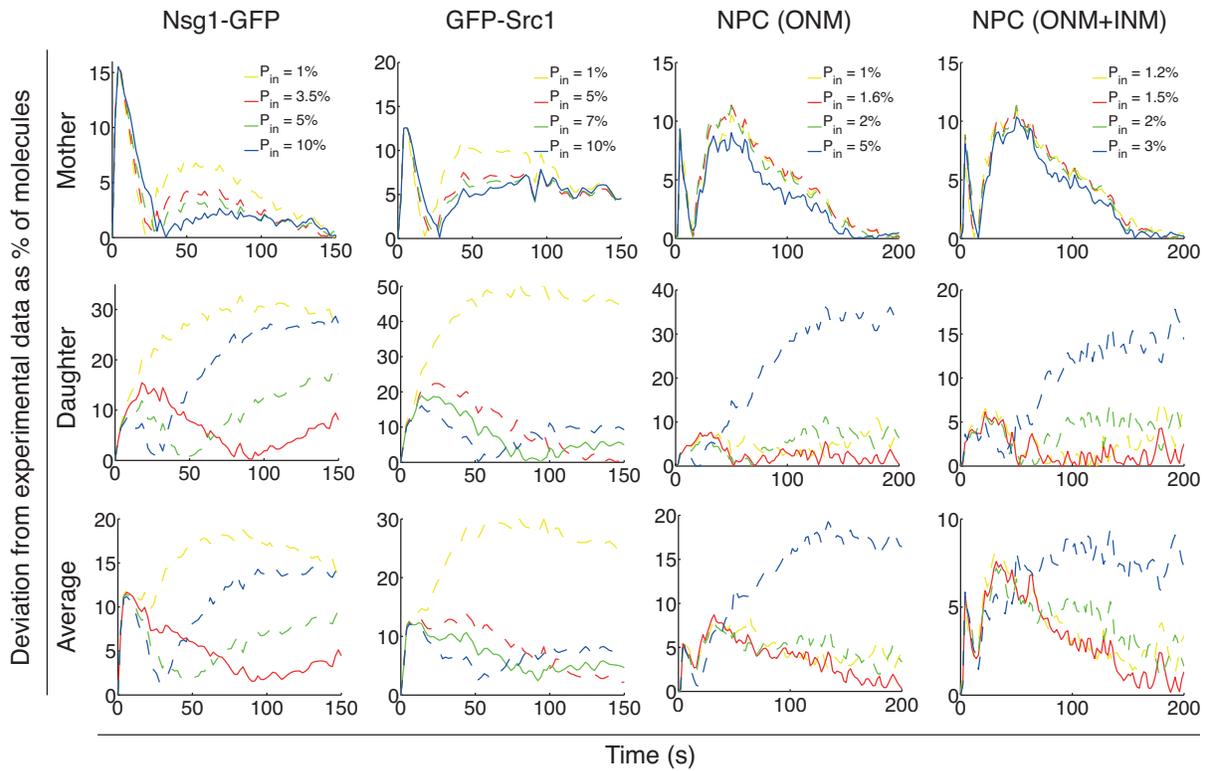

**FIGURE S6. Estimated $P_{in}$ values for the specialised lipid ring domain at the neck of EA nuclei.** Deviations (in percentages) of stochastic simulations from the experimental mean, for each experimental time step. Mother and daughter lobe deviations were calculated as the absolute value of the difference between simulations and FLIP experiments. The average deviation is the mean of both nuclear lobes' deviations over time. The best fit (i.e. smallest deviation) is shown as a continuous line. Data from Nsg1-GFP, GFP-Src1 and Nup49-GFP FLIP profiles in EA was used (21 cells). For the NPC data, we considered both scenarios where the specialised lipid domain lies only at the ONM or at the whole NE (ONM + INM). Effective diffusion coefficients were fixed as in Table S2 and we let $P_{out}$ = 100 % in all cases. The estimated $P_{in}$ values were $P_{in}$ = 3.5 % for Nsg1-GFP (ONM), $P_{in}$ = 7 % for GFP-Src1 (INM), $P_{in}$ = 1.6 % for NPCs (domain at ONM) and $P_{in}$ = 1.5 % for NPCs (domain at ONM and INM).

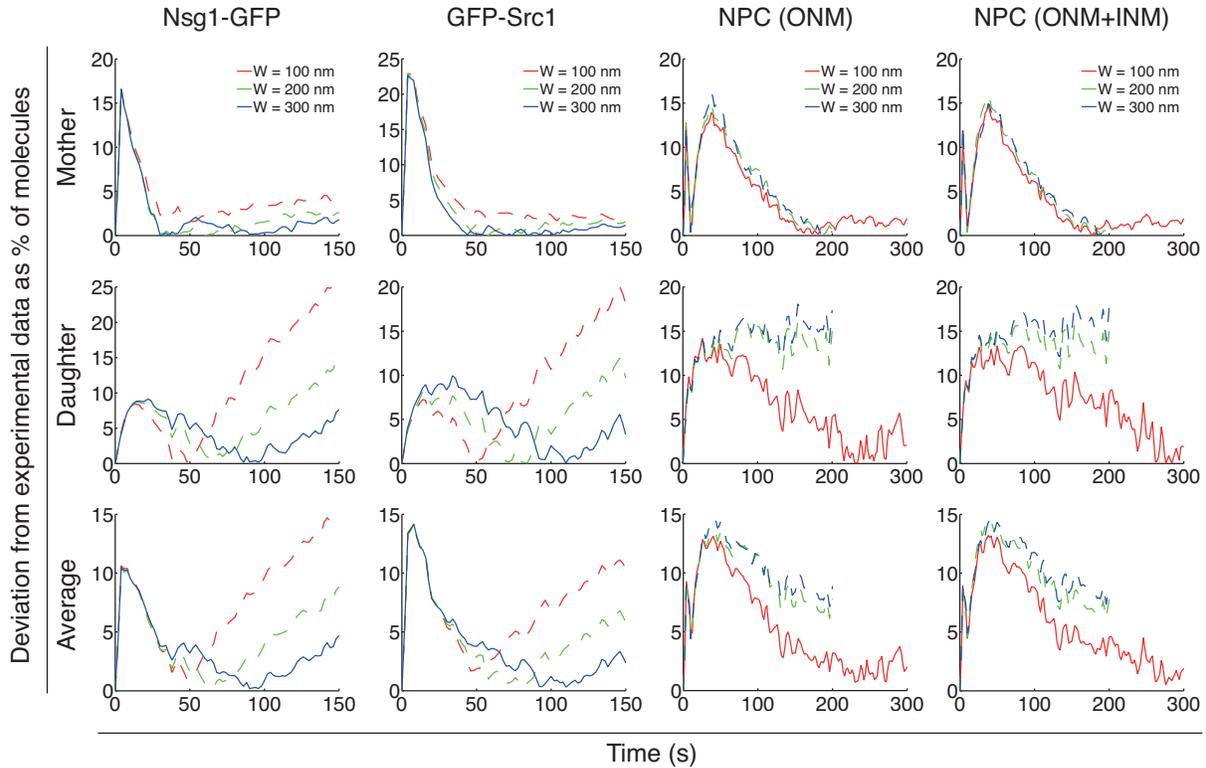

**FIGURE S7. Estimated width of a single specialised lipid ring domain centred at the bridge of LA nuclei.** Deviations (in percentages) of stochastic simulations from the experimental mean, for each experimental time step. Mother and daughter lobe deviations were calculated as the absolute value of the difference between simulations and FLIP experiments. The average deviation is the mean of both nuclear lobes' deviations over time. The best fit (i.e. smallest deviation) is shown as a continuous line. Data from Nsg1-GFP, GFP-Src1 and Nup49-GFP FLIP profiles in LA was used (34 cells). For the NPC data, we considered both scenarios where the domain lies only at the ONM or at the whole NE (ONM + INM). Effective diffusion coefficients were fixed as in Table S2 while $P_{in}$ values were also fixed as in Table 1 (LA, one ring centred at the bridge). As before, we let $P_{out}$ = 100 % in all cases. The estimated width of the specialised single ring domain was of 300 nm for Nsg1-GFP (ONM) and GFP-Src1 (INM), while a width of 100 nm fitted best for NPCs, independently of whether the domain was assumed only at the ONM or at the whole NE (ONM + INM).

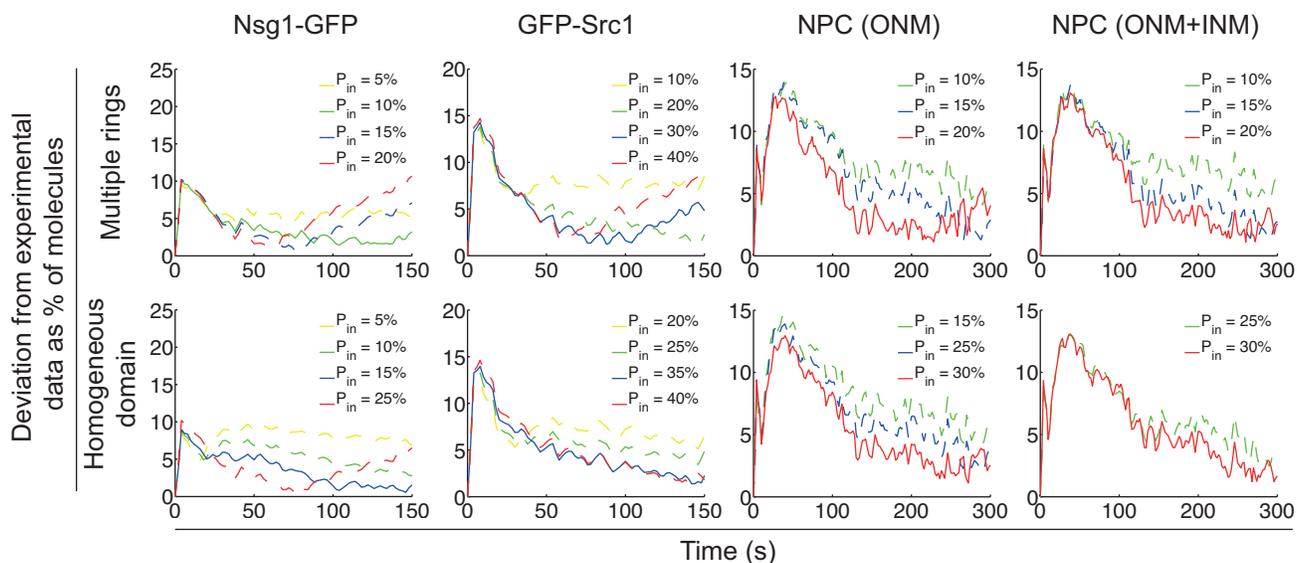

**FIGURE S8. Estimated $P_{in}$ values for two different specialised lipid domain configurations at the bridge of LA nuclei.** Average deviations (in percentages) of stochastic simulations from the experimental mean for each experimental time step. Mother and daughter lobe deviations are calculated as the absolute value of the difference between simulations and FLIP experiments. The average deviation is the mean of both nuclear lobes' deviations over time. The best fit (i.e. smallest deviation) is shown as a continuous line. For the NPC, we considered both scenarios where the specialised lipid domain lies only at the ONM or at both ONM and INM. Effective diffusion coefficients were fixed as in Table S2 and we fixed $P_{out}$ = 100 % in all cases. The estimated $P_{in}$ values are shown in Table 1.

**TABLE S1. Diffusing protein parameters for spatial-stochastic simulations.** Protein numbers were taken from the Yeast GFP Fusion Localization Database [11,52]. Source references for protein sizes are indicated. Sizes used in simulations take into account GFP fusion. An exception is the NPC, given that Nup49 is buried in its inner rings.

| Molecular species | Initial number | Diffusion rate | Estimated size (including GFP) | Reference |
|---|---|---|---|---|
| GFP | - | - | (W x L) 2.4 x 4.2 nm | [1] |
| TetR-GFP | 5000 @ nucleoplasm | 1.9 $\mu m^2$/s | 4 nm (8 nm) | [2] |
| NPC (Nup49-GFP) | 150 @ periplasm | 0.2 $\mu m^2$/s | 98 nm | [3] |
| Nsg1-GFP | 1900 @ ONM | 0.3 $\mu m^2$/s | ~ 4.2 nm (8 nm) | [4] |
| GFP-Src1 | 2100 @ INM | 0.3 $\mu m^2$/s | ~ 6 nm (10 nm) | [4] |

**TABLE S2. Estimated parameters of sphingolipid domains.** Nuclear membranes estimated thickness, viscosity, viscous drag, diffusion coefficient and permeability of the sphingolipid domain. Values outside and inside of the sphingolipid domain are specified where applies. All values were estimated using the Petrov-Schwille model.

| Parameter | Description | Value | Value outside of sphingolipid domain | Value inside of sphingolipid domain |
|---|---|---|---|---|
| $h_{ONM}$, $h_{INM}$ | Membrane thickness | - | 4 nm | 5.4 nm |
| $h_p$ | Perinuclear space thickness at connecting bridge | 13 ± 4 nm | 13 nm | If domain only at ONM: 12.3 nm<br>If domain at ONM and INM: 11.6 nm |
| $\mu_c$, $\mu_n$, $\mu_p$ | Bulk viscosities of cytoplasm, nucleoplasm and periplasm | 11.967 × 10$^{-4}$ Pa s | - | - |
| $\eta_{ONM}$ | Surface viscosity of ONM | - | 7.5333 × 10$^{-9}$ Pa s m | 2.6826 × 10$^{-8}$ Pa s m |
| $\eta_{INM}$ | Surface viscosity of INM | - | 7.2424 × 10$^{-9}$ Pa s m | 2.5976 × 10$^{-8}$ Pa s m |
| $\eta_{NE}$ | Surface viscosity of NE<br>$\eta_{ONM} + \eta_{INM} + \mu_p h_p$ | - | 1.4791 × 10$^{-8}$ Pa s m | If domain only at ONM: 3.4083 × 10$^{-8}$ Pa s m<br>If domain at ONM and INM: 5.2816 × 10$^{-8}$ Pa s m |
| $\gamma_{Nsg1}$ | Viscous drag experienced by Nsg1-GFP | - | 1.3951 × 10$^{-8}$ Pa s m | 4.1853 × 10$^{-8}$ Pa s m |
| $\gamma_{Src1}$ | Viscous drag experienced by GFP-Src1 | - | 1.3951 × 10$^{-8}$ Pa s m | 4.1853 × 10$^{-8}$ Pa s m |
| $\gamma_{NPC}$ | Viscous drag experienced by NPC | - | 2.0926 × 10$^{-8}$ Pa s m | If domain only at ONM: 7.3943 × 10$^{-8}$ Pa s m<br>If domain at ONM and INM: 1.0655 × 10$^{-7}$ Pa s m |
| $D_{Nsg1-GFP}$ | Effective diffusion rate of Nsg1-GFP | - | 0.3 μm$^2$/s | 0.1 μm$^2$/s |
| $D_{GFP-Src1}$ | Effective diffusion rate of GFP-Src1 | - | 0.3 μm$^2$/s | 0.1 μm$^2$/s |
| $D_{NPC}$ | Effective diffusion rate of NPC | - | 0.2 μm$^2$/s | Domain only at ONM: 0.0566 μm$^2$/s<br>Domain at ONM and INM: 0.0393 μm$^2$/s |